\documentstyle[prd,aps,psfig]{revtex}
\bibstyle{unsrt}

\tighten
\begin{document}
\draft

\preprint{SUSX-TH-97-012,HD-THEP-97-33}
\twocolumn[\hsize\textwidth\columnwidth\hsize\csname 
@twocolumnfalse\endcsname

\title{The Thermodynamics of Cosmic String densities in $U(1)$ 
Scalar Field Theory}
\author{Nuno D. Antunes$^1$, Lu\'{\i}s M. A.  Bettencourt$^2$ and 
Mark Hindmarsh$^1$}
\address{$^1$School of Mathematical and Physical Sciences,
University of Sussex, Brighton BN1 9QH, U.K.}
\address{$^2$Institut f\"ur Theoretische Physik, Universit\"at Heidelberg,
Philosophenweg 16, 69120 Heidelberg, Germany}
\date{\today}
\maketitle

\begin{abstract}
We present a full characterization of the phase transition in 
U(1) scalar field theory and of the associated vortex string 
thermodynamics in 3D. 
We show that phase transitions in the string densities exist and 
measure their critical exponents, both for the long string and the short 
loops. Evidence for a natural separation between these two string 
populations is  presented. In particular our results strongly indicate 
that an infinite string population will only exist above the critical 
temperature. Canonical initial conditions for cosmic string evolution 
are show to correspond to the infinite temperature limit of the theory.
\end{abstract}

\pacs{PACS Numbers : 05.70.Fh, 11.27.+d, 98.80.Cq \hfill   HD-THEP-97-33, 
SUSX-TH-97-012 }

\vskip2pc]

Topological defects appear in a great variety of systems from condensed 
matter laboratory experiments to the early Universe.
Their importance in phase transitions in the laboratory 
is know to be fundamental and their presence in the early Universe 
may be the key to many of the unsolved questions in standard cosmology.

However, in spite of the universal relevance of topological defects, 
much about the fundamental description of their formation and evolution 
remains qualitative. This is a reflection of the complexities involved  
in  first principle studies, owing to their nature as non-perturbative 
excitations of quantum field theories.

Particularly interesting for cosmology are string-like topological defects
\cite{CS}. 
Motivated by the study of their creation in the early Universe 
\cite{ZurekN}, a variety of experiments 
has been recently developed with the aim of studying 
vortex string formation in liquid crystals \cite{liqcrys}, 
superfluid  $^4$He \cite{Helium4} 
and $^3$He \cite{Helium3} systems.
Their new results permit us to test with unprecedented precision 
theoretical ideas about defect formation and evolution. 

From a theoretical standpoint, cosmic strings and other topological defects, 
have been traditionally thought to be produced at phase transitions in the 
early Universe as a result of the formation of correlated domains 
\cite{Kibble}.  
More recently a refinement of this scenario \cite{Zurek} has been gaining 
support, that most topological defects existing below the phase transition 
are, in fact, survivors of a population of unstable 
defects existing above the phase transition as non-linear excitations 
of the fields.
In order  to exist at low energies these field configurations 
need to be  frozen in by some non-equilibrium process, 
which is context dependent. In  
a continuous phase transition, like the 
superfluid transition in $^4$He, this process may in addition be 
helped by the critical slowing down in the response of the fields  
close to the critical point.
An estimate of the defect density hence produced is  usually achieved given 
the correlation length of the field at the time of freeze out. 
By assigning random phases to different domains and connecting by the path 
of minimum phase gradient, a network of defects can in principle be 
constructed.
This picture generates order of magnitude estimates 
that successfully predicted the defect densities observed 
in the Helium experiments. 

In practice higher correlations in the phase structure of the fields exist and 
can change the picture considerably.  As an illustration of the shortcomings
of simply using the correlation length to predict produced defect densities 
note that  similar densities would be estimated if the defect network 
were frozen in above or below the critical point, provided that the 
correlation length were chosen to be the same in both cases. 
This is clearly not what is observed experimentally \cite{Helium4},
suggesting that the string network must be very different in two 
circumstances were the domain structure can be expected to coincide.   

In order to match theory to future, more accurate measurements, as well 
as for the sake of theoretical understanding it is highly desirable to 
generate more detailed predictions for the string density formed 
at phase transitions. 
In cosmology and  in the laboratory,  equilibrium 
is the natural starting point. If we know the statistics of strings at any 
given  temperature the reference to the domain structure in the 
fields  becomes obsolete, while the role of non-equilibrium  
in freezing in the string network, at least in some scales, can 
be expected to hold as usual. 

In this Letter we present the full characterization of the behavior of the 
string densities with temperature in the U(1) scalar field theory. 
In doing so we go  
beyond existing studies in the XY (see eg. \cite{XY})
model, which belongs to the same 
universality class, especially in determining the behavior of infinite strings,
measuring critical exponents associated with string densities 
and confronting our findings  with cosmological scenarios of defect 
formation.
  
We first determine the field thermodynamics and locate the critical 
point. In order to drive the system to equilibrium we evolved the fields  
stochastically according to
\begin{eqnarray}
\left( \partial_{t}^2 -\nabla^2 \right) \phi_i - m^2 \phi_i
+ \lambda \phi_i \left( \sum_j \phi_j^2 \right) 
+  \eta \dot{\phi_i} = \xi_i, 
\label{e1}
\end{eqnarray}
where $\{i,j\} \in {1,2}$. The stochastic variable
$\xi_i({\bf x},t)$ is taken to be  Gaussian, characterized by 
\begin{eqnarray}
& \langle & \xi_i({\bf x},t) \rangle = 0, \\
& \langle & \xi_i({\bf x},t) \xi_j({\bf x'},t')\rangle= 
  {2 \eta \over{\beta}} \delta({\bf x-x'}) \delta(t-t') \delta_{i j}.
\label{e2}
\end{eqnarray}
These relations ensure that detailed balance applies 
in equilibrium, which will result for large times at temperature $1/\beta$.
We evolved the system Eq.~(\ref{e1}-\ref{e2}) 
using a staggered-leapfrog method, 
on a lattice of size $100^3$. We chose the lattice spacing to be 
$\delta l = 0.5$. We verified 
that the correlation length was always resolved by at least 4 points in each 
linear dimension, apart from the cases when $\beta \ll 1$. 
Around the critical point all physical scales become much larger than the 
lattice spacing. Universal quantities can then be measured and shown 
explicitly to be  independent of this ultraviolet cutoff. 
Away from the critical point, at high or low temperatures,
universality is lost in general. There, the results loose some of their 
physical meaning but retain model properties, that are interesting from a 
theoretical point of view. This will be the case of the infinite temperature 
limit of the theory.    

An equivalent equilibrium state could have been obtained using lattice 
Monte Carlo methods. Also, from the equilibrium point of view, the 
second order 
time derivative is redundant. However, this work is intended to be 
the precursor 
to non-equilibrium studies of the relativistic field theory, 
where is is more natural to use a Langevin 
method such as ours, and the field equations are naturally 
second order in time derivatives. 

\begin{figure}
\centerline{\psfig{file=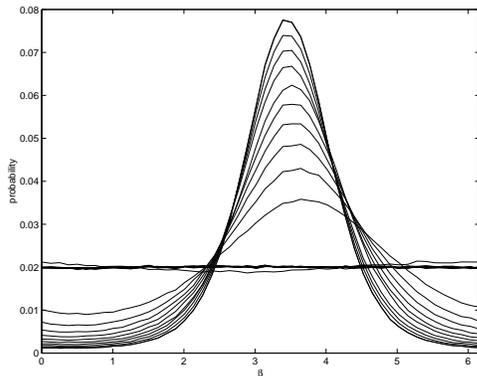,width=2.5in}}
\caption{Phase probability distributions around the critical temperature
 for $\beta$ between 1.1 and 2.9 (steps of 0.1).
 For low temperatures the distribution becomes 
peaked around a randomly chosen phase. As the temperature increases 
the distribution 
becomes flat and the $U(1)$ symmetry is restored.}
\label{fig1}
\end{figure}

We decide when the system has reached equilibrium by monitoring 
several quantities associated with different length scales. We measure the 
kinetic energy of the system and check for equipartition. This characterizes
the smallest scales in the sample. We monitor the string densities, 
associated with intermediate length scales  and we measure the phase 
distribution functions across the whole volume.
By making sure that  these three quantities have reached asymptotic 
average values we conclude that the system is in equilibrium. 
The temperature of the bath  coincides with that measured using 
equipartition.

The transition between a system displaying a prefered random direction in its 
manifold of field minima at low temperature and a flat phase distribution, 
characteristic of the $U(1)$ symmetry is shown in Fig.~\ref{fig1}.

\begin{figure}
\centerline{\psfig{file=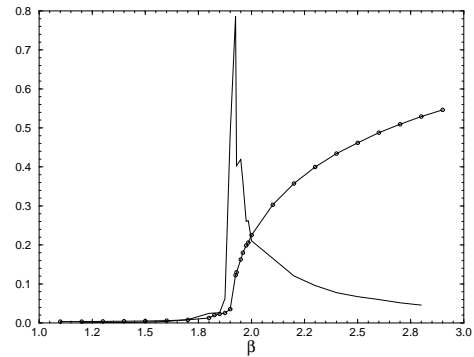,width=3.0in,angle=270}}
\caption{The  variation of $\langle \vert \bar{\phi}\vert \rangle$ 
and its derivative, rescaled by $1/4$, 
with the temperature. Circles denote measured points. 
The divergence of the derivative marks the critical point.}
\label{fig2}
\end{figure}

Although appealing the phase probability plots of Fig.~\ref{fig1} do not 
permit a precise determination of the critical temperature or any of its 
associated critical exponents.
The critical point can be more easily found by observing the variation
of the modulus of the spatial average of the field with the temperature:
\begin{eqnarray}
\langle \vert \bar{\phi}\vert \rangle =\left\langle \sqrt{\sum_{i=1,2} \left( 
{1\over{V}}\int_V d{\bf x}~ \phi_{i}({\bf x}) \right)^{2} }\right
\rangle,
\label{e3}
\end{eqnarray}
where the angular brackets denote ensemble averaging, obtained in practice 
by averaging over many independent measurements, $200-1000$, 
corresponding to well separated evolution times. 
The variation of Eq.~(\ref{e3}) and its derivative 
with temperature is shown in Fig.~\ref{fig2}. 
In particular we see that the derivative diverges
at the critical point.

The critical temperature  can be measured by assuming 
 that Eq.~(\ref{e3}) behaves like
\begin{eqnarray}
\langle \vert \bar\phi(\beta) \vert \rangle = A\;(\beta -\beta_c)^\alpha,
\label{e4}
\end{eqnarray}
which is assumed to hold just below the critical temperature. 
By fitting our data at several temperatures to Eq.~(\ref{e4}), for different 
choices of the inverse critical temperature $\beta_c$, 
we can determine its minimal $\chi^2$ value. 
Simultaneously we obtain the critical exponent $\alpha$.
The critical values hence determined are $\beta_c=1.906 \pm 0.008$ and
$\alpha=0.43 \pm 0.07$. The universal value of 
$\alpha$ is in reasonable agreement with renormalization group calculations 
\cite{Zinn}. 

Having measured the critical behavior in the fields we want 
to analyze how vortex string densities vary with the temperature. 
We assume ergodicity of the field evolution once equilibrium is reached.
Using this fact we analyze at given time intervals the phase of the complex 
field, and associate a vortex to each 2-D lattice cell where the $U(1)$ phase 
winds through $2 \pi$. We then proceed to connect the vortices  and 
construct the string network.
Given a string network, we measure the string density in loops and long 
string, for different values of the bath temperature around the phase 
transition.  

\begin{figure}
\centerline{\psfig{file=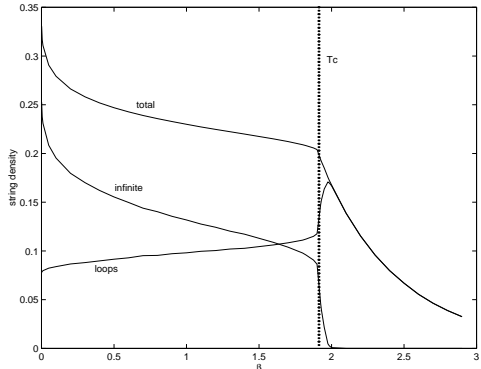,width=2.5in}}
\caption{The lattice link densities of total string 
$\rho_{tot}$, loops $\rho_{l}$
 and infinite string $\rho_{i}$ as a function of inverse temperature $\beta$ .}
\label{fig3}
\end{figure}

Fig.~\ref{fig3} shows how the total density per lattice link of strings
 $\rho_{tot}$, as well as the density of long and short string ($\rho_{i}$ and 
 $\rho_{l}$ respectively) changes with temperature.
These quantities allow us a direct comparison to algorithms of cosmic string 
formation.

We observe a dramatic change in the behavior of the  string system 
across the phase transition. The curves of 
fig.~\ref{fig3} suggest that we can decompose 
strings into two distinct populations, one of loops, say strings smaller 
than a certain cutoff length $L_\infty$ 
of the order of the squared linear dimension of our 
total volume  and another population consisting of much longer strings. 
We will refer to the former as the string loop population and to 
the latter, owing to the usual terminology in cosmology, as infinite 
strings.  

 At $\beta_c$ all three densities display
abrupt changes in their behavior. These changes are  phase transitions 
as in all three cases the derivatives of the densities present discontinuities
at the critical point.
  Even more remarkable is the fact that the appearance
of infinite string seems to be linked to the criticality in the fields.
The very sharp rise in the infinite string density  at the critical
point may suggest that the phase transition associated with it may 
actually be discontinuous in the infinite volume limit. 
 
\begin{figure}
\centerline{\psfig{file=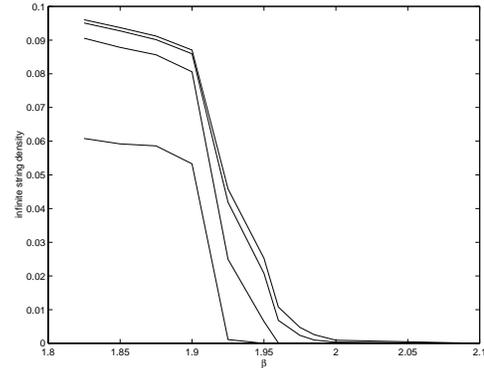,width=2.5in}}
\caption{The dependence of $\rho_i$ on inverse Temperature, 
for $L_\infty = 2500, 8000, 20000, 100000 $ (top to bottom), measured in 
lattice units.}
\label{fig4}
\end{figure}

In order to clarify this question we measured the infinite string density 
variation with $\beta$ for various values of $L_\infty$. 
The results are shown in Fig.~\ref{fig4}.
The temperature at which infinite strings first appear 
clearly depends on the choice of $L_\infty$ , getting closer to $\beta _c$
as it increases. We also verified that for smaller lattice sizes
infinite string first appeared at lower temperatures, eg. in a $20^3$
computational domain there was infinite string as soon as $\beta=2.3$.
Both these variations supply evidence that as the volume is increased the 
temperature at which infinite strings first appear migrates towards the 
critical point implying a discontinuous phase transition in $\rho_{i}$.
At present it is however impossible to perform such extrapolation
with full confidence.

We can establish that the phase transition 
in the fields and in the string densities coincide  by finding the 
critical point and exponents for the latter. 

At the critical temperature  the density of strings drops suddenly and keeps
falling for lower temperatures. The decay of 
 $\rho_{tot}$   
which coincides with  $\rho_{l}$, is excellently 
described by the  exponential law 
\begin{eqnarray}
\rho_{tot}(\beta) = \rho_{\rm tot}(\beta_c) e^{-(\beta -\beta_c) E}
\label{e5}
\end{eqnarray}
where $\rho_{tot}(\beta_c) \simeq 0.2$ is a universal number 
(see eg. \cite{XY} for a very different study yielding the same result) 
and $E=1.836 $, 
implying that at temperatures below the critical point strings are 
Boltzmann suppressed.
At low temperatures (in practice for $\beta \gtrsim 6$) there is long 
range order and no strings survive. 

Just above the critical temperature the variation of the string densities 
can be best characterized by defining the quantity 
$ \delta \rho \equiv (\rho(T) - \rho (T_c)) /\rho (T_c)$. 
The behavior of the infinite string density $\rho_{ i}$ 
is well described by an ansatz similar to Eq.~(\ref{e4}), 
\begin{eqnarray}
\delta \rho_{ i} \propto (T-T_c)^{\gamma_{ i}}.
\label{e7}
\end{eqnarray}
Note the change from $\beta$ in Eq.~(\ref{e4}) to $T$ in Eq.~(\ref{e7}). 
We find $\gamma_{ i } = 0.25 \pm 0.01$ and 
$\beta_c = 1.911 \pm 0.001$, in agreement with all previous estimates.
For $\rho_{ tot}$ we find:
\begin{eqnarray}
\delta \rho_{tot} \propto ( T - T_c)^{\gamma_{tot}},
\label{e8}
\end{eqnarray}
with $\beta_c = 1.912 \pm 0.002$ and ${\gamma_{ tot}} = 0.39 \pm 0.01$.

For temperatures well above the phase transition the loop density 
can be reasonably described by a quadratic form
\begin{eqnarray}
\rho_{ l}(\beta) 
\simeq \rho_{l}(0) \left( 1 +  {\beta \over 3} - {1 \over 3} 
({\beta \over 2} )^2 \right) 
\label{e9}
\end{eqnarray}
where  $\rho_{l}(0)\simeq 1/12$. This corresponds to a fraction of 
$25  \pm 1 \%$ of all strings in loops for $\beta=0$.

The variations are much more complicated for $\rho_{ tot}$ 
and $\rho_{ i}$. 
The large size of the derivatives close to 
$\beta=0$ make it hard to approximate the curves and a  successful fit 
can be achieved by a polynomial of very large degree, typically 9 or 10, 
with coefficients of alternating sign. Indeed, the behavior looks 
non-polynomial instead, which would mean that there is no 
standard high-temperature expansion for the infinite string density. 
This is explicable if we realize that there is no local operator which
can distinguish an infinite string, and thus we would not
necessarily expect a polynomial behavior around $\beta=0$.

Particularly interesting is the infinite temperature limit.
We observe that then the total string density per link tends to 
$\rho_{tot}(0)=1/3$ and that its fraction of long strings tends 
to $75 \pm 1 \% $.  
These results coincide with those obtained from the canonical 
Vachaspati-Vilenkin (VV) algorithms for simulating  string formation 
in the early Universe \cite{VV,Andy,AchBorLid97}.  
Their underlying motivation results from the picture that strings should
result from random phase fluctuations within spatial domains of a given 
size. Typically, the complex field is given unit modulus, with phases assigned 
randomly to each site on a cubic lattice. There are variations on this theme: 
for example, the continuous phase may be approximated by three
values \cite{VV}, which leads to a slightly different string density and 
infinite string fraction. Nonetheless, assigning phases randomly to each 
lattice site corresponds precisely to the infinite temperature limit, 
and thus VV algorithms generate ensembles of string at infinite temperature. 
These ensembles are used as initial conditions for the free evolution 
of the string network, using a numerical solution to the classical
dynamics of relativistic strings \cite{CS}.  
Thus numerical simulations of this 
type model an instantaneous quench from infinite to zero temperature.
While these initial conditions are  ultimately unphysical, 
the network of strings is observed to 
approach rapidly a self-similar evolution, which serves to disguise the
initial state. 
On the other hand, if one would like to describe a network 
of strings below the critical point the Vachaspati-Vilenkin algorithm would be 
a completely inappropriate starting point. 

Finally, our observation that the phase transition in the fields may
coincide with the point when infinite string starts forming, suggests that
strings do indeed play a fundamental role in the critical behavior of
the system, as has been suggested so often in the literature 
\cite{Onsager,Feynman,Williams,Kleinert}. 
On the other hand, statistical studies of string networks \cite{EdRay} 
commonly identify the critical point in string densities 
with the Ginzburg temperature $T_G < T_c$. %(assuming mean-field exponents 
%and our value of $\beta_c$, $\beta_G= 2.156$) 
Above we presented evidence 
for the coincidence of the critical behavior in the fields and in 
the strings  at $T_c$.  In a future publication \cite{Us},
a closer look at the properties of the string length distribution 
and a more detailed scale dependence study will help to clarify these problems.
%$\beta_G = 2.156$  

We thank the Department of Computer Science at the Technical University 
of Berlin and the Fujitsu/Imperial College Centre for Parallel 
Computing for generous allocation of supercomputer time. 
We thank Andy Yates for usage of his string tracing routine. 
NDA thanks Ed Copeland for useful comments. 
LMAB is supported by the {\it Deutsches Forschung Gemeinschaft}.  
NDA is supported by JNICT - {\it Programa Praxis} XXI,
under contract BD/2794/93-RM. MH is supported by PPARC (UK), under 
grants B/93/AF/1642,  GR/L12899, and GR/K55967. 
This work is partially supported by the European Science Foundation.

\end{document}